# On the Zagreb Indices Equality

**Hosam Abdo$^a$, Darko Dimitrov$^a$, Ivan Gutman$^b$**

$^a$*Institut für Informatik, Freie Universität Berlin,*
*Takustraße 9, D–14195 Berlin, Germany*
E-mail: `[abdo,darko]@mi.fu-berlin.de`

$^b$*Faculty of Science, University of Kragujevac,*
*P. O. Box 60, 34000 Kragujevac, Serbia*
E-mail: `gutman@kg.ac.rs`

**Abstract**

For a simple graph $G$ with $n$ vertices and $m$ edges, the first Zagreb index and the second Zagreb index are defined as $M_1(G) = \sum_{v \in V} d(v)^2$ and $M_2(G) = \sum_{uv \in E} d(u)d(v)$. In [34], it was shown that if a connected graph $G$ has maximal degree 4, then $G$ satisfies $M_1(G)/n = M_2(G)/m$ (also known as the Zagreb indices equality) if and only if $G$ is regular or biregular of class 1 (a biregular graph whose no two vertices of same degree are adjacent). There, it was also shown that there exist infinitely many connected graphs of maximal degree $\Delta = 5$ that are neither regular nor biregular of class 1 which satisfy the Zagreb indices equality. Here, we generalize that result by showing that there exist infinitely many connected graphs of maximal degree $\Delta \geq 5$ that are neither regular nor biregular graphs of class 1 which satisfy the Zagreb indices equality. We also consider when the above equality holds when the degrees of vertices of a given graph are in a prescribed interval of integers.

## 1 Introduction

Let $G = (V, E)$ be a simple graph with $n = |V|$ vertices and $m = |E|$ edges. For $v \in V$, $d(v)$ is its degree. The *first Zagreb index* $M_1(G)$ and the *second Zagreb index* $M_2(G)$ are defined as follows:

$$M_1(G) = \sum_{v \in V} d(v)^2 \quad \text{and} \quad M_2(G) = \sum_{uv \in E} d(u)d(v).$$

For the sake of simplicity, we often use $M_1$ and $M_2$ instead of $M_1(G)$ and $M_2(G)$, respectively.

In 1972 the quantities $M_1$ and $M_2$ were found to occur within certain approximate expressions for the total $\pi$-electron energy [16]. In 1975 these graph invariants were proposed to be measures of branching of the carbon-atom skeleton [15]. The name "Zagreb index" (or, more precisely, "Zagreb group index") seems to be first used in the review article [4]. For details of the mathematical theory and chemical applications of the Zagreb indices see surveys [10, 14, 25, 30] and papers [12, 13, 36, 37, 38].

We denote by $K_{a,b}$ the *complete bipartite* graph with $a$ vertices in one class and $b$ vertices in the other one. Let $D(G)$ be the set of the vertex degrees of $G$, i.e., $D(G) = \{d(v) \,|\, v \in V\}$. The *subdivision* graph $S(G)$ of a graph $G$ is obtained by inserting a new vertex (of degree 2) on every edge of $G$. A *regular* graph is a graph where each vertex has the same degree. A regular graph with vertices of degree $k$ is called a $k$-regular graph.

The graph $G$ is *biregular* if its vertex degrees assume exactly two distinct values. We distinguish between two types of biregular graphs: biregular graphs of *class* 1 have the property that no two vertices of the same degree are adjacent. In biregular graphs of *class* 2 at least one edge connects vertices of equal degree.

Let $G$ be a graph with $n$ vertices and let $a$, $b$, and $c$ be three positive integers, $1 \leq a \leq b \leq c \leq n-1$. The graph $G$ is said to be *triregular* if for $i = 1, 2, \ldots, n$, either $d_i = a$ or $d_i = b$ or $d_i = c$, and there exists at least one vertex of degree $a$, at least one vertex of degree $b$, and at least one vertex of degree $c$. If so, then $G$ is a triregular graph of degrees $a$, $b$, and $c$, or for brevity, an $(a, b, c)$-triregular graph. Similarly, as in the case of biregular graphs, we distinguish two types of triregular graphs: Triregular graphs of *class* 1 have the property that no two vertices of the same degree are adjacent. In triregular graphs of *class* 2 at least one edge connects vertices of equal degree.

As defined in [1], a set $S$ of integers is *good* if for every graph $G$ with $D(G) \subseteq S$, the inequality (1) holds. Otherwise, $S$ is a *bad* set.

## 1.1 Comparing Zagreb indices

In spite of the fact that the two Zagreb indices were introduced simultaneously and examined almost always together, relations between them were not considered until quite recently. Observe that, for general graphs, the order of magnitude of $M_1$ is $O(n^3)$ while the order of magnitude of $M_2$ is $O(mn^2)$. This suggests comparing $M_1/n$ and $M_2/m$ instead of $M_1$ and $M_2$. Based on his *AutoGraphiX* [6] conjecture-generating computer system, Pierre Hansen arrived at the inequality

$$\frac{M_1(G)}{n} \leq \frac{M_2(G)}{m} \tag{1}$$

which he conjectured to hold for all connected graphs. In the current mathematico-chemical literature, the relation (1) is usually referred to as the *Zagreb indices inequality*. If the equality case is excluded, then we speak of the *strict Zagreb indices inequality*. Soon after the announcement of this conjecture it was shown [18] that there exist graphs for which (1) does not hold. Although the work [18] appeared to completely settle Hansen's conjecture, it was just the beginning of a long series of studies [1, 2, 5, 8, 19, 21, 23, 26, 27, 32, 33] in which the validity or non-validity of either [18] or some generalized version of [18] was considered for various classes of graphs. These studies are summarized in two recent surveys [23, 24]. We briefly mention some known results.

The inequality (1) holds for trees [32], unicyclic graphs [31], and graphs of maximum degree four, so called molecular graphs [18], graphs with only two distinct vertex degrees.

In [1] it was shown that the Zagreb indices inequality holds for graphs with vertex degrees in the set $\{s - c, s, s + c\}$, for any integers $c$, $s$. This implies that the inequality holds for graphs with vertex degrees from any interval of length three. Sun and Chen [26] proved that any graph $G$ with maximum vertex dgrees $\Delta(G)$ and minimun vertex degrees $\delta(G)$, such that $\Delta(G) - \delta(G) \leq 3$ and $\delta(G) \neq 2$ satisfy (1). Thus, any interval $[x, x+3]$ is good with only exception of $[2, 5]$. In [1], this result was enhanced by showing that the inequality holds for graphs with vertex degrees from an interval $[c, c + \lceil \sqrt{c}\, \rceil]$ for any integer $c$. Therefore, if $G$ is a graph with $\Delta(G) - \delta(G) \leq \lceil \sqrt{c}\, \rceil$ and $\delta(G) \geq c$ for some integer $c$, then $G$ satisfies the inequality (1). It also imples that there are arbitrary long good intervals.

The last result was strengthened in [2], where it was proved that for every positive integer $p$, the interval $[a, a+p]$ is good if and only if $a \geq p(p-1)/2$ or $[a, a+p] = [1, 4]$. In [2] also, an algorithm for deciding if a given set of integers $S$ of cardinality $s$ is good, which requires $O(s^2 \log s)$ time and $O(s)$ space was presented.

Recently, in [34] it was shown that the Zagreb indices inequality (1) holds for the subdivision graph $S(G)$ of any graph $G$, biregular graphs of class 1 (strict inequality holds for biregular graphs of class 2), $(a, b, c)$-triregular graph of class 1 (strict inequality holds for connected $(a, b, c)$-triregular graph of class 2), union of complete graphs from distinct cardinalities greater than 1, union of $p$-complete graph and $q$-cycle graph for all $p \leq 1$, $q \geq 3$, union of $p$-complete graph and $q$-path graph, $q \geq 3$ for all $p$, $q$ (strict inequality), union of $p$-cycle graph and $q$-path graph for all $p$, $q$ (strict inequality), union of $p$-path graph and $q$-path graph for all $p$, $q$, and the union of $p$-cycle graph and complete bipartite graph $K_{a,b}$, $a \leq b$ for all $p$, $a$, $b$ except for $p \geq 3$, $a = 1$, $b \geq 5$.

On the other side there are graphs that do not satisfy the inequality (1), even more, there is an infinite family of planar graphs of maximum degree $\Delta \geq 5$ such that the inequality (1) is false [1]. See [1, 18, 19, 32] for various examples of graphs dissatisfying this inequality. In [8, 18, 19, 26, 27, 32], examples of connected simple graph $G$ are given such that $M_1/n > M_2/m$.

Curiously, however, in spite of such an extensive research on inequality (1), little attention was paid on the equality case, i.e., on the characterization of graphs for which

$$\frac{M_1(G)}{n} = \frac{M_2(G)}{m} \tag{2}$$

holds. In the line with above notation, we call (2) the *Zagreb indices equality*.

To prove some of the results in this paper, we exploit a decomposition of $M_2/m - M_1/n$ introduced by Hansen and Vukičević [18]. Denote by $m_{i,j}$ the number of edges that connect vertices of degrees $i$, $j$ in the graph $G$, then

$$\begin{aligned}\frac{M_2}{m} - \frac{M_1}{n} &= \frac{\sum_{v \in V} d(v)^2}{m} - \frac{\sum_{uv \in E} d(u)d(v)}{n} \\ &= \sum_{\substack{i \leq j, k \leq l \\ (i,j),(k,l) \in \mathbb{N}^2}} \left[ \left( ij \left( \frac{1}{k} + \frac{1}{l} \right) + kl \left( \frac{1}{i} + \frac{1}{j} \right) - i - j - k - l \right) m_{i,j}\, m_{k,l} \right]. \end{aligned} \tag{3}$$

Further analyzing of (3) can be simplified by introducing the function

$$f(i, j, k, l) = ij \left( \frac{1}{k} + \frac{1}{l} \right) + kl \left( \frac{1}{i} + \frac{1}{j} \right) - i - j - k - l,$$

with variables $i, j, k, l \in \mathbb{N}$, and studying its properties. Now, (3) can be restated as

$$\frac{M_2}{m} - \frac{M_1}{n} = \sum_{\substack{i \leq j \\ k \leq l \\ (i,j),(k,l) \in \mathbb{N}^2}} f(i, j, k, l) m_{i,j}\, m_{k,l}.$$

Notice that the function $f$ can be represented in the following way

$$f(i, j, k, l) = (ij - kl) \left( \frac{1}{k} + \frac{1}{l} - \frac{1}{i} - \frac{1}{j} \right) = (ij - kl) \frac{ij(k + l) - kl(i + j)}{ijkl}. \tag{4}$$

Some properties of the function $f$ have been studied in [1, 2].

Easy verification shows that the Zagreb indices equality holds for regular graphs and stars. In [34] it was shown that the Zagreb indices equality holds for the subdivision graph $S(G)$ of $r$-regular graph and $r > 0$, union of complete graphs that have same cardinality, union of $p$-complete graph and $q$-cycle graph for $p = 3$, $q \geq 3$, union of $p$-path graph and $q$-path graph for $p = q = 2$, and $p = q = 3$, union of $p$-cycle graph and complete bipartite graph $K_{a,b}$, $a \leq b$ only for $p \geq 3$, $a = b = 2$ and $p \geq 3$, $a = 1$, $b = 4$.

Also, as in [34], it was shown that if a connected graph $G$ has maximal degree 4, then $G$ satisfies the Zagreb indices equality if and only if $G$ is regular or biregular of class 1. There, it was also shown that there exist infinitely many connected graphs of maximal degree $\Delta = 5$ that are neither regular nor biregular of class 2, which satisfy the Zagreb indices equality. The example used there was a $(a, b, c)$-triregular of class 2. In the next section, we generalize that result by showing that there exist infinitely many connected graphs of maximal degree $\Delta \geq 5$ that are neither regular nor biregular of class 1, which satisfy the Zagreb indices equality. In Section 3, we characterize when the above equality holds when the degrees of vertices of a given graph are in the prescribed intervals of integers.

## 2 Connected graphs of maximal degree $\Delta \geq 5$

**Theorem 2.1.** *There exist infinitely many connected graphs $G$ of maximum degree $\Delta \geq 5$ that are neither regular nor biregular of class 1 that satisfy the Zagreb indices equality.*

*Proof.* Consider the connected graph $G(x, y, z, w)$ depicted in Figure 1. The graph $G(x, y, z, w)$ is based on $x$ copies of $K_{2,5}$, one copy of $K_{2,z}$ and $w$ copies of $K_{3,3}$. The construction of $G(x, y, z, w)$ is as follows:

- Make a sequence of $x$ copies of $K_{2,5}$. Let us denote the edges of $K_{2,5}^i$ by $u_1^i v_1^i, u_1^i v_2^i, \ldots, u_1^i v_5^i, u_2^i v_1^i, u_2^i v_2^i, \ldots, u_2^i v_5^i$ for all $1 \leq i \leq x$. The connection between two consecutive copies of $K_{2,5}$ is founded by replacing the edges $u_2^i v_5^i$ and $u_1^{i+1} v_1^{i+1}$ by the edges $u_2^i v_1^{i+1}$ and $u_1^{i+1} v_5^i$ respectively. Continue this kind of replacement between all consecutive copies of $K_{2,5}$. Notice that these replacements do not change the degrees of the vertices.

- Next, denote the vertices of $K_{2,z}$ with degree $z$ by $t_1$ and $t_2$, and the vertices with degree two by $p_1, p_2, \ldots, p_z$. Remove the edges $t_2 p_1$ and $t_1 p_z$. Connect a path on $2y$ vertices with the vertex $v_5^x$ and the vertex $p_1$ and a vertex of degree two with the vertex $u_2^x$ and the vertex $t_1$. These replacements also do not change the degrees of the vertices.

- Next, insert two adjacent vertices $t$ and $s$. Connect $t_2$ with $t$, $p_z$ with $s$, and $t$ with $s$.

- Make a sequence of $w$ copies of $K_{3,3}$. Denote the vertices of $K_{3,3}^i$ by $a_1^i, a_2^i, a_3^i$, and $b_1^i, b_2^i, b_3^i$. Replace the edge $a_1^i b_1^i$ by the path $a_1^i a^i a_3^i$ and $a_3^i b_3^i$ by the path $b_1^i b^i b_3^i$. Connect $s$ with $a^1$. Further, connect $b^i$ with $a^{i+1}$, for $i = 1, \ldots, w - 1$. Finally, insert a vertex $q$ adjacent to $b^w$, $u_1^1$, and $v_1^1$. Notice that all vertices are of degree 3.

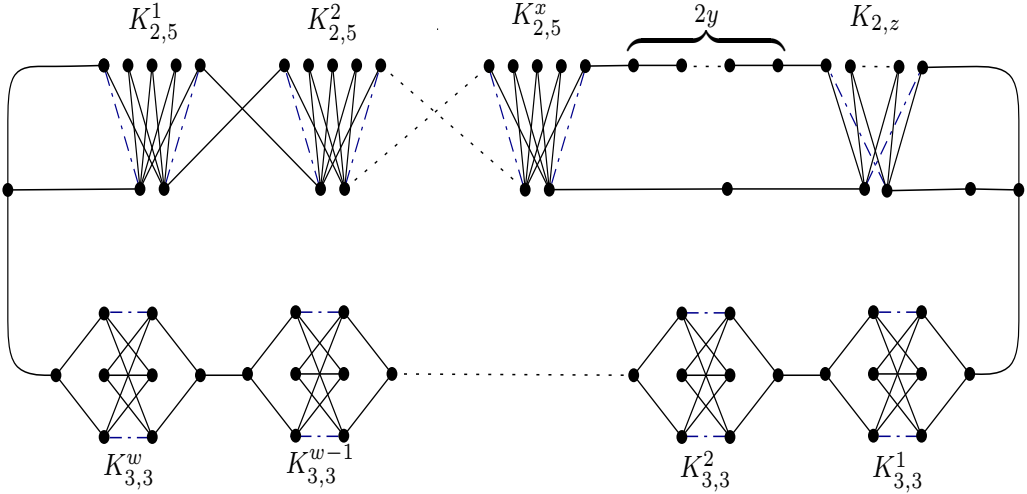

**Figure 1:** A connected graph $G(x, y, z, w)$ based on $x$ copies of $K_{2,5}$, one copy of $K_{2,z}$, and $w$ copies of $K_{3,3}$. The dash-dotted edges are those that are removed from the corresponding complete bipartite graphs.

The graph $G(x, y, z, w)$ has $2x$ vertices of degree 5, $8w + 2$ of degree 3, $5x + 2y + z + 2$ vertices of degree 2 and two vertices of degree $z$. The values of the positive $m_{i,j}$, $i, j \in \mathbb{N}$, are: $m_{z,2} = 2z$, $m_{5,2} = 10x - 1$, $m_{3,2} = 3$, $m_{2,2} = 2y+1$, $m_{5,3} = 1$, and $m_{3,3} = 12w+1$. Then $n = 7x+2y+z+8w+6$, $m = 10x+2y+2z+12w+5$, $M_1 = 2(35x+4y+z^2+2z+36w+13)$, and $M_2 = 100x+8y+4z^2+108w+36$. The graph $G(x, y, z, w)$ satisfies the Zagreb indices equality if

$$\begin{aligned} mM_1 - nM_2 &= -86 - 242x - 28y + 36z - 264w - 36xy + 80xz + 4xw + 16yz - 40yw + 84zw - \\ & \quad 8xz^2 - 4yz^2 - 8wz^2 - 6z^2 \\ &= 0. \end{aligned}$$

From here, we have that the expression $mM_1 - nM_2$ equals zero if there are $x, y, z, w \in \mathbb{N}$ that satisfy

$$x = \frac{132w - 42zw + 4z^2w + 14y - 8yz + 20yw + 2yz^2 + 3z^2 - 18z + 43}{-121 - 18y + 2w + 40z - 4z^2}. \tag{5}$$

For any $x, y, z, w \in \mathbb{N}$, it holds that $132w - 42zw + 4z^2w > 0$, $14y - 8yz + 20yw + 2yz^2 > 0$ and $3z^2 - 18z + 43 > 0$. Therefore the nominator in (5) is also positive. The denominator in (5) equals 1 if

$$w = 61 + 9y - 20z + 2z^2. \tag{6}$$

For any $y, z \in \mathbb{N}$ there exist $w \in \mathbb{N}$ such that (6) holds. Thus, for an arbitrary value of $z$, one can obtain infinitely many instances of $G(x, y, z, w)$ that satisfy the Zagreb indices equality. □

## 3  Graphs with vertex-degrees from prescribed intervals

In this section, we consider the case when the degrees of vertices of a given graph are in a prescribed interval of integers. In [2], it was shown that if the vertex degrees of an $n$-vertex graph $G$ are from the interval $[a, a+p]$, $a \geq p(p-1)/2$ where $p$ is a positive integer not exceeding $\lfloor \frac{1}{2}(\sqrt{8n-7}-1) \rfloor$, then $G$ satisfies the Zagreb indices inequality. Here, we prove that, except very few cases (see Theorem 3.1), the graphs with vertex degrees from the interval $[a, a+p]$, $a \geq p(p-1)/2$, $p \in \mathbb{N}$ do not satisfy the Zagreb indices equality. To show that result, we analyze the equation (4), more precisely, we investigate when $f(i, j, k, l) = 0$., i.e., when

$$ij = kl \tag{7}$$

or

$$\frac{i+j}{ij} = \frac{k+l}{kl}. \tag{8}$$

First, we consider the equality (7).

**Lemma 3.1.** *There are no two different pairs of integers $x, y$ and $u, v$ from an interval $[a, a+p]$, $a \geq \dfrac{p(p-1)}{2}$, $a, p \in \mathbb{N}$, that satisfy $xy = uv$.*

*Proof.* Assume that there are different pairs $x, y$ and $u, v$ that satisfy $xy = uv$. We also assume that $x < u \leq v < y$ and $x = p(p-1)/2 + p_1 + k$, $y = p(p-1)/2 + p_4 + k$, $u = p(p-1)/2 + p_2 + k$, and $v = p(p-1)/2 + p_3 + k$ where $0 \leq p_1 < p_2 \leq p_3 < p_4 \leq p$, and $k$ is nonnegative integer. The variable $k$ determines the offset of the beginning of the interval $[a, a+p]$ from $p(p-1)/2$. Now, $xy = uv$ can be restated as

$$\left(\frac{p(p-1)}{2} + p_1 + k\right)\left(\frac{p(p-1)}{2} + p_4 + k\right) = \left(\frac{p(p-1)}{2} + p_2 + k\right)\left(\frac{p(p-1)}{2} + p_3 + k\right),$$

or
$$\frac{p(p-1)}{2}(p_1 + p_4 - p_2 - p_3) = (p_2 + k)(p_3 + k) - (p_1 + k)(p_4 + k). \tag{9}$$

We prove that (9) cannot be fulfilled, and therefore the assumption that there are different pairs $x, y$ and $u, v$ that satisfy $xy = uv$ is false. So, we prove
$$\frac{p(p-1)}{2}(p_1 + p_4 - p_2 - p_3) \neq (p_2 + k)(p_3 + k) - (p_1 + k)(p_4 + k). \tag{10}$$

First, we prove the lemma for $k = 0$, i.e., $x, y, u, v \in [p(p-1)/2, (p+1)p/2]$, by showing that
$$p_1 p_4 > p_2 p_3 \quad \text{if and only if} \quad p_1 + p_4 \geq p_2 + p_3. \tag{11}$$

Let $p_2 = p - 1 + c_1$ and $p_3 = p_4 - c_2$, where $c_1, c_2 \in \mathbb{N}$ and $c_1, c_2 < p$.

To prove the "if" direction of (11), we show that if $p_1 + p_4 < p_2 + p_3$ then $p_1 p_4 \leq p_2 p_3$. From $p_1 + p_4 < p_2 + p_3$, we have then $p_1 + p_4 < p_1 + c_1 + p_4 - c_2$ and $c_1 > c_2$. Now,
$$\begin{aligned}
p_1 p_4 - p_2 p_3 &= p_1 p_4 - (p_1 + c_1)(p_4 - c_2) \\
&= c_1 c_2 - p_4 c_1 + p_1 c_2 \\
&= c_1 (c_2 - p_4) + c_2 p_1 \\
&= c_2 p_1 - c_1 p_3 \\
&< 0.
\end{aligned}$$

To prove the other direction of (11), we show that if $p_1 p_4 \leq p_2 p_3$ then $p_1 + p_4 < p_2 + p_3$. Indeed,
$$\begin{aligned}
p_1 p_4 - p_2 p_3 \leq 0 &\Rightarrow -p_4 c_1 + p_1 c_2 + c_1 c_2 \leq 0 \\
&\Rightarrow -p_3 c_1 + p_1 c_2 \leq 0 \\
&\Rightarrow p_1 c_2 \leq p_3 c_1 \\
&\Rightarrow c_2 < c_1 \\
&\Rightarrow p_4 - p_3 < p_2 - p_1 \\
&\Rightarrow p_1 + p_4 < p_2 + p_3.
\end{aligned}$$

To complete the proof of the lemma we show that (10) holds for $k \geq 1$, by showing that when $p_1 + p_4 - p_2 - p_3$ is non negative, $(p_2 + k)(p_3 + k) - (p_1 + k)(p_4 + k)$ is negative, and vice versa. First, if $p_1 p_4 > p_2 p_3$ by (11), it follows that $p_1 + p_4 \geq p_2 + p_3$. Then,
$$\begin{aligned}
p_1 + p_4 \geq p_2 + p_3 &\Rightarrow k(p_2 + p_3) \leq k(p_1 + p_4) \\
&\Rightarrow p_2 p_3 + k(p_2 + p_3) + k^2 \leq p_1 p_4 + k(p_1 + p_4) + k^2 \\
&\Rightarrow (p_2 + k)(p_3 + k) < (p_1 + k)(p_4 + k).
\end{aligned}$$

Second, if $p_1 p_4 \leq p_2 p_3$ by (11), it follows that $p_1 + p_4 < p_2 + p_3$. Then,
$$\begin{aligned}
p_1 + p_4 < p_2 + p_3 &\Rightarrow k(p_2 + p_3) > k(p_1 + p_4) \\
&\Rightarrow p_2 p_3 + k(p_2 + p_3) + k^2 > p_1 p_4 + k(p_1 + p_4) + k^2 \\
&\Rightarrow (p_2 + k)(p_3 + k) > (p_1 + k)(p_4 + k).
\end{aligned}$$
□

Next, we investigate when (8) is fulfilled. The main characterization is given in Lemma 3.2. Before we present it, we need the following three propositions.

**Proposition 3.1.** *Let the integers $u, v$ belong to an interval $[x, y]$, where $x = \frac{p(p-1)}{2}$, $y = x + p$, and $p \in \mathbb{N}$. Then $uv > xy$ if and only if $u + v \geq x + y$.*

*Proof.* We assume that $v \leq u$. Let $u = x + p_1$ and $v = x + p_2$, where $p_1, p_2 \in \mathbb{N}$ and $p_1 \geq p_2$.

First prove that if $uv > xy$ then $u + v \geq x + y$, which is equivalent to show that if $u + v < x + y$ than $uv \leq xy$. We prove the last implication. Now,

$$u + v = 2x + p_1 + p_2, \quad x + y = 2x + p, \quad \text{and} \quad u + v < x + y \implies p_1 + p_2 < p.$$

Next,

$$uv = (x + p_1)(x + p_2) = x^2 + x(p_1 + p_2) + p_1 p_2.$$

With the constrain $p_1 + p_2 < p$, the last expression has its maximum for $p_1 = p_2 = \dfrac{p-1}{2}$. Thus

$$uv \leq x^2 + x(p_1 + p_2) + \frac{(p-1)^2}{4}. \tag{12}$$

On the other hand

$$xy = x(x + p) = x^2 + xp \geq x^2 + x(p_1 + p_2 + 1) = x^2 + x(p_1 + p_2) + x. \tag{13}$$

Since, $x = p(p-1)/2 > (p-1)^2/4$, from (12) and (13) we have $uv < xy$.

Now, we prove that if $x + y \leq u + v$ then $xy < uv$. From $x + y \leq u + v$, we have that $p \leq p_1 + p_2$. Next

$$xy = x(x + p) = x^2 + xp \leq x^2 + x(p_1 + p_2), \tag{14}$$

and

$$uv = (x + p_1)(x + p_2) = x^2 + x(p_1 + p_2) + p_1 p_2. \tag{15}$$

From (14) and (15), together with $p_1 \geq p_2 > 0$, it follows that $xy < uv$. $\square$

The following proposition shows that if the positive integers $u, v$ are from interval $[x, y]$, $x \geq p(p-1)/2$, $y = x + p$, then expression $(u + v)/uv = (x + y)/xy$ can be satisfy only if $x = p(p-1)/2$ and $y = x + p$.

**Proposition 3.2.** *Let the integers $u, v$ belong to an interval $[x, y]$, where $x = \dfrac{p(p-1)}{2}$, $y = x + p$, $p \in \mathbb{N}$ and $u', v' \in (x', y')$, where $x' = x + k$, $y' = y + k$, $u' = u + k$, $v' = v + k$, $k$ being positive integer. Then, $\dfrac{x' + y'}{x'y'} \neq \dfrac{u' + v'}{u'v'}$.*

*Proof.* We assume that $u \leq v$. Let $u = x + p_1$, $v = x + p_2$. Then, we have $p_1 \leq p_2$. Let

$$\begin{aligned} g(x, y, u, v) &= (x + y)uv - (u + v)xy \\ &= uvx + uvy - uxy - vxy \\ &= (x + p_1)(x + p_2)(2x + p) - (2x + p_1 + p_2)xy \\ &= x^2(p_1 + p_2 - p) + p_1 p_2 (x + y), \end{aligned}$$

and

$$\begin{aligned} g(x', y', u', v') &= (x' + y')u'v' - (u' + v')x'y' \\ &= k^2(u + v - x - y) + 2k(uv - xy) + g(x, y, u, v). \end{aligned} \tag{16}$$

The inequality $\dfrac{x' + y'}{x'y'} \neq \dfrac{u' + v'}{u'v'}$ holds if and only if $g(x', y', u', v') \neq 0$.

First, consider the case $xy < uv$. By Proposition 3.1, it follows that $x + y \leq u + v$. Thus, for the first two terms of (16), we have, $k^2(u + v - x - y) \geq 0$ and $2k(uv - xy) > 0$. Also, from $x + y \leq u + v$, we have $p \leq p_1 + p_2$. This implies $g(x, y, u, v) = x^2(p_1 + p_2 - p) + p_1 p_2(x + y) > 0$ and finally $g(x', y', u', v') > 0$.

Second, consider the case $xy > uv$. By Proposition 3.1, it follows that $x + y > u + v$. Thus $k^2(u + v - x - y) < 0$ and $2k(uv - xy) < 0$. Also from $x + y < u + v$, we have $p_1 + p_2 < p$. With this constraint, the function $g(x, y, u, v)$ attains its maximum at $p_1 = p_2 = (p-1)/2$. Therefore, $g(x, y, u, v) \leq -x^2 + (p-1)^2(2x+p)/4$. Further substituting $x$ by $(p-1)p/2$, we obtain $g(x, y, u, v) \leq 0$, and finally $g(x', y', u', v') < 0$.

Notice that the case $xy = uv$ by Lemma 3.1 is not possible. □

In the next proposition we show that if two different pairs $x, y$ ($x \leq y$) and $u, v$ ($u \leq v$), $x \leq u$, from $[a, a + p]$, $a \geq p(p-1)/2$ satisfy (8) then $x = a$, $y = a + p$.

**Proposition 3.3.** *Let the integers $x, y, u, v$ belong to an interval $[a, a + p]$, $a \geq \dfrac{p(p-1)}{2}$, $p \in \mathbb{N}$ and $x \leq y$, $u \leq v$, $x \leq u$. If $\dfrac{x+y}{xy} = \dfrac{u+v}{uv}$ then $x = a$ and $y = a + p$.*

*Proof.* We prove the proposition by induction on $p$. For $p = 1$ and $p = 2$ an easy verification shows that there are no two different pairs of integers $x, y$ and $u, v$ that satisfy $(x + y)/xy = (u + v)/uv$. For $p = 3$ the only 4-tuple that satisfies $(x + y)/xy = (u + v)/uv$ is 3, 6, 4, 4 [26]. Assume that the claim is true for $p$.

Now consider the intervals of length $p + 1$. By the induction hypothesis, if there are pairs $x, y$ and $u, v$ from an interval of length $p$ that satisfy $(x + y)/xy = (u + v)/uv$, then $x = a$ and $y = a + p$, where $a \geq p(p-1)/2$. By Proposition 3.2, the interval $[p(p-1)/2, p(p-1)/2 + p]$ is the only interval of length $p$ for which $(x + y)/xy = (u + v)/uv$, where $x = p(p - 1)/2$, $y = p(p - 1)/2 + p$, and $u, v \in [p(p - 1)/2, p(p - 1)/2 + p]$. Let $I_{p+1} = [a_{p+1}, a_{p+1} + p + 1]$ be an interval of length $p + 1$. It holds that $a_{p+1} \geq p(p-1)/2 + p$. Thus, all subintervals of length $p$, [x,y], of an interval of length $p+1$ does not satisfy $(x+y)/xy = (u+v)/uv$. If there is a 4-tuple $x, y, u, v$ from an interval of length $p+1$ that satisfies $(x + y)/xy = (u + v)/uv$, two of these elements must be $a_{p+1}$ and $a_{p+1} + p + 1$. Assume that it is not true that $x = a_{p+1}$ and $y = a_{p+1} + p + 1$. Then, $x \leq u \leq y \leq v$ or $x \leq y \leq u \leq v$. In all these cases it is easy to verify that $(x + y)/xy \neq (u + v)/uv$. □

Finally, we characterize for which pairs from an interval $[a, a + p]$, $a \geq p(p-1)/2$, $p \in \mathbb{N}$, the equation (8) is fulfilled.

**Lemma 3.2.** *Let the integers $x, y, u, v$ belong to an interval $[a, a+p]$, $a \geq \dfrac{p(p-1)}{2}$, $p \in \mathbb{N}$, such that $x \leq y$ and $u \leq v$. Then, $\dfrac{x+y}{xy} = \dfrac{u+v}{uv}$ if $p$ is odd and $x = \dfrac{p(p-1)}{2}$, $y = x + p$ and $u = v = x + \dfrac{p-1}{2}$.*

*Proof.* If there are such integers $x, y, u, v$ that satisfy $(x+y)/xy = (u+v)/uv$, then by Proposition 3.3 $x = p(p - 1)/2$ and $y = p(p - 1)/2 + p$. Let $u = x + h$, $v = y - k$, $h, k \in \mathbb{N}$. The equation $(x + y)/xy = (u + v)/uv$ is satisfies if and only if $h(x, y, u, v) = 0$. Substituting $u$ and $v$ in $h(x, y, u, v)$, we have

$$\begin{aligned} h(x,y,u,v) &= (u+v)xy - (x+y)uv \\ &= (x+y+h-k)xy - (x+y)(x+h)(y-k) \\ &= \frac{p^2}{4}(k - h + 4hk - 2p(k+h) + p^2(k-h)). \end{aligned} \qquad (17)$$

First, consider the case when $k > h$.

Let $p$ be even. From $u \leq v$ and $k > h$, it follows that $h = 1, \ldots, p/2 - 1$, $k = h, \ldots, p - 1$. The expression (17), has extreme point (saddle point) at $(h, k) = (-(p - 1)^2/4, (p + 1)^2/4)$ which lie outside the valid range of $h$ and $k$. The minimum value of $h(x, y, u, v)$ in the valid range of $h$ and $k$ ($0 < h, k < p$) is bigger than 0, and it is obtained at $h = p/2 - 1$ and $k = p/2$. So for even $p$ we have $h(x, y, u, v) > 0$.

If $p$ is odd, then the minimum value of $h(x, y, u, v)$ in the valid range of $h$ and $k$ is equal 0, at $h = (p - 1)/2$ and $k = (p + 1)/2$.

Second, consider the case $k < h$. Then the expression (17), has its minimum value bigger than 0, in the valid range of $h$ and $k$, at $h = (p - 1)/2$ and $k = (p + 1)/2$.

If $k = h$, then $h(x, y, u, v) = p^2(4hk - 2p(k + h))/4 = p^2(4h^2 - 4ph)/4$. Since $h < p$, we have $h(x, y, u, v) < 0$.

Thus, we conclude that $(x + y)/xy = (u + v)/uv$ only when $p$ is odd and $u = v = (p^2 - 1)/2$. □

Now, we are ready to determine the graphs with vertex degrees from a prescribed interval (dis)satisfying the Zagreb indices equality.

**Theorem 3.1.** *Let $G$ be a graph with $D(G) \subseteq [a, a + p]$, $a \geq \dfrac{p(p-1)}{2}$, or $D(G) \subseteq [1, 4]$. Then, $G$ satisfies the Zagreb indices equality if*

(a) *$G$ is regular graph,*

(b) *$G$ is biregular graph of class 1,*

(c) *$G$ is disjoint union of $\dfrac{(p-1)(p+1)}{2}$-regular graphs and biregular graph of class 1 with degree of vertices $\dfrac{(p-1)p}{2}$ and $\dfrac{p(p+1)}{2}$, where $p$ is odd, or*

(d) *$G$ is disjoint union of stars $S_5$ and cycles of arbitrary length.*

*Proof.* In [2], it was shown that $f(i, j, k, l) \geq 0$ whenever $i, j, k, l \in [a, a + p]$ and $a \geq p(p - 1)/2$. Consequently, if $G$ fulfills the Zagreb indices equality, then all $f(i, j, k, l)$ must equal zero.

First, let $|D(G)| = 1$, with $D(G) = \{a\}$. Then $G$ is regular, and fulfills the Zagreb index equality since $f(a, a, a, a) = 0$.

Second, let $|D(G)| = 2$, with $D(G) = \{a, b\}$. Since $f(a, a, b, b) > 0$, if $G$ satisfies the Zagreb indices equality, $G$ does not contains edges with endvertices of same degrees. Thus $G$ must contains only edges with endvertices $ab$, i.e., $G$ is biregular graph of class 1.

Now consider the case $|D(G)| \geq 3$. Recall that by (4) $f(x, y, u, v) = 0$ if $xy = uv$ or $(x + y)/xy = (u + v)/uv$. By Lemma 3.1 there are no two different pairs $(x, y)$ and $(u, v)$ such that $xy = uv$. By Lemma 3.2, $(x + y)/xy = (u + v)/uv$ is fulfilled only if $p$ is odd and $x = p(p - 1)/2$, $y = p(p + 1)/2$ and $u = v = (p + 1)(p - 1)/2$. Thus only in those cases $f(x, y, u, v) = 0$. The resulting graph $G$ that satisfy the Zagreb indices equality must be a disjoint union of $((p - 1)(p + 1)/2)$-regular graphs and biregular graph of class 1 with degree of vertices $p(p - 1)/2$ and $p(p + 1)/2$, where $p$ is odd.

It is easy to verify that the only pairs from the interval $[1, 4]$ that satisfy the Zagreb indices equality are the pairs $1, 4$ and $2, 2$. In this case $G$ must be a disjoint union of stars $S_5$ and cycles of arbitrary length. □

As an immediate consequence of Theorem 2.1, we have the following corollary.

**Corollary 3.1.** *Let $G$ be a connected graph with $D(G) \subseteq [a, a + p]$, $a \geq \dfrac{p(p-1)}{2}$, or $D(G) \subseteq [1, 4]$. If $|D(G)| > 2$ then, $G$ does not satisfies the Zagreb indices equality.*

By Theorem 2.1 we have the next corollary.

**Corollary 3.2.** *Let $I$ be an interval such that $I = [a, a + p]$, $a \geq \dfrac{p(p-1)}{2}$, or $I = [1, 4]$. Then, there exist infinitely many graphs $G$ with $D(G) \nsubseteq I$, such that $G$ satisfies the Zagreb indices equality.*

Notice that by Theorem 2.1, a graph $G$ that satisfy Corollary 3.2 has $D(G) = \{2, 3, 5, a\}$, $a \in I$. We believe that a strengthened version of Corollary 3.2 also holds.

**Conjecture 3.1.** *Let $I$ be an interval such that $I = [a, a+p]$, $a \geq \dfrac{p(p-1)}{2}$, or $I = [1, 4]$. Then, for any other interval $I_n \not\subseteq I$ there exist infinitely many graphs $G$ with $D(G) \subseteq I_n$ such that $G$ satisfies the Zagreb indices equality.*